\documentclass[nofootinbib]{revtex4}
\usepackage{graphicx}
\usepackage{latexsym}
\def\be{\begin{equation}}
\def\ee{\end{equation}}
\def\bea{\begin{eqnarray}}
\def\eea{\end{eqnarray}}

\begin{document}

\title{ Spontaneous Leptogenesis in Brans-Dicke Cosmology}
\author{Chi-Yi Chen${}^{a,b}$}
\email{chenchiyi@center.shao.ac.cn}

\author{You-Gen Shen${}^{a,c,e}$}
\email{ygshen@center.shao.ac.cn}

\author{Bo Feng${}^{b,d}$}
\email{fengbo@mail.ihep.ac.cn}

%\author{Mingzhe Li${}^d$}
%\email{limz@mail.ihep.ac.cn}

%\author{Xinmin Zhang${}^d$}
%\email{xmzhang@mail.ihep.ac.cn}

\affiliation{${}^a$Shanghai Astronomical Observatory, Chinese
Academy of Sciences, Shanghai 200030, PR China}
\affiliation{${}^b$Graduate School of Chinese Academy of Sciences,
Beijing 100049, PR China}
 \affiliation{${}^c$National Astronomical
Observatories, Chinese Academy of Sciences, Beijing 100012, PR
China} \affiliation{${}^d$Institute of High Energy Physics, Chinese
Academy of Science, P.O. Box 918-4, Beijing 100049, PR China}
\affiliation{${}^e$Institute of Theoretical Physics, Chinese Academy
of Sciences, Beijing 100080, PR China.}

\begin{abstract}
The role of the auxiliary scalar field $\phi$ of Brans-Dicke
theory played in baryon number asymmetry is discussed in this
paper. We consider a derivative coupling of this gravitational
scalar field to the baryon current ${J^{\mu}}_B$ or the current of
the baryon number minus lepton number ${J^{\mu}}_{B-L}$ based on a
series of works of R. Morganstern about the Brans-Dicke cosmology.
We find that the spontaneous baryogenesis by this coupling is
capable to yield a sufficient baryon asymmetry $n_B/s\sim 10^{10}
$ for the time of the grand unification is in a little advanced.
In addition, Davoudiasl et al have recently introduced a new type
of interaction between the Ricci scalar $R$ and the bayon current
$J^{\mu}$, $\partial_{\mu}R J^{\mu}$ and also proposed a mechanism
for baryogenesis, the gravitational baryogenesis. However, the
Einstein equation tell us that $\dot{R}=0$ in the
radiation-dominated epoch of the standard FRW cosmology. In this
paper we reconsider the feasibility of having gravitational
baryongenesis with such a form of interaction in radiation-filled
Brans-Dicke cosmology. We will show that $\dot{R}$ does not vanish
in this case and the required baryon number asymmetry can also
be achieved. \\
PACS number(s): 98.80.Cq, 11.30.Er
\end{abstract}

\maketitle

\section{introduction}

 \hspace*{3.5mm}The origin of the baryon number
asymmetry remains a big puzzle in cosmology and particle physics.
Conventionally, it is argued that this asymmetry is generated from
an initial baryon symmetry phase dynamically as long as the
following conditions are satisfied \cite{sakharov}: (1)baryon
number non-conserving interactions; (2) $C$ and $CP$ violations;
(3)out of thermal equilibrium. When the $CPT$ is violated
dynamically, however the baryon number asymmetry can be generated
in thermal equilibrium \cite{cohen}. In connecting to dark energy,
a class of models of spontaneous leptogenesis
\cite{zhang,li,trodden}are investigated recently by introducing a
interaction between the dynamical dark energy scalars and the
ordinary matter. Specifically, a derivative coupling of the
quintessence scalar field $Q$ to the baryon or lepton current is
under considering:
\begin {eqnarray}
{\cal{L}}_{int} \sim \partial_{\mu} Q J^{\mu}.
\end {eqnarray}
One silent feature of this scenario for baryogenesis is that the
present accelerating expansion and the generation of the matter
and antimatter asymmetry of our universe is described in a unified
way.

\hspace*{3.5mm}Recently, a new mechanism called gravitational
baryogenesis in thermal equilibrium has been proposed by
Davoudiasl et al \cite{davoudiasl}. They introduced explicitly an
interaction between the Ricci scalar curvature with derivative and
the baryon number current:
\begin {eqnarray}
{\cal{L}} = \frac{1}{M^2} \partial_{\mu}R J^{\mu}.
\end {eqnarray}
And the baryon number asymmetry is given by
\begin {eqnarray}
\frac{n_B}{s}\sim \frac{\dot{R}}{M^2T},
\end {eqnarray}
which shows that $n_B/s$ is determined by the value of $\dot{R}$,
however the Einstein equation, $R=8\pi G{T^{\mu}}_{\mu}=8\pi
G(1-3\omega)\rho$, tells us that $\dot{R}=0$ in the
radiation-dominated epoch of the standard
Friedmann-Robertson-Walker (FRW) cosmology. Davoudiasl et al in
Ref.\cite{davoudiasl} have considered three different
possibilities of obtaining a non-vanishing $\dot{R}$ which include
the effects of trace anomaly, reheating and introducing a
non-thermal component with $\omega>1/3$ dominant in the early
universe. In the brane-world scenario Shiromizu and Koyama in
Ref.\cite{shiromizu} provided another example for $\dot{R}$
example for $\dot{R}\neq0$. A generalized form of the derivative
coupling of the Ricci scalar to the ordinary matter
\begin {eqnarray}
{\cal{L}}_{int}\sim \partial_{\mu} f(R)J^{\mu}.
\end {eqnarray}
is also discussed in \cite{hong}. Taking $f(R)\sim \ln R$, they
have shown that $\partial_{\mu}f(R)\sim \partial_{\mu}R/R$ does
not vanish and the required baryon asymmetry can also be generated
in the early universe.

As it is well known, Brans-Dicke theory of gravity is one of the
simplest extensions/modifications of Einstein's general
relativity\cite{bd,b,scalar}. Compared with general relativity, as
well as the metric tensor of space-time which describes the
geometry there is an auxiliary scalar field $\phi$ which also
describes the gravity. The role of the gravitational scalar in
Brans-Dicke theory has been much exploited to achieve inflation,
quintessence
\cite{Yoshimura,McDonald,Sakai,Sen,Carroll,Perrotta,Chiba,Mak,Trodden}
and k-essence \cite{Kim}. The main constraints on this
gravitational scalar come from the solar system
experiment\cite{Will}. In addition, the testing of Brans-Dicke
theory using stellar distances, the CMB temperature and
polarization anisotropy have also been discussed in
\cite{gaztanaga,xueleichen}. In this regard, here we focus
particularly on the scalar of the Brans-Dicke theory to model
baryo(lepto)genesis behavior by combining both the spontaneous
symmetry breaking and its gravitational origin in Brans-Dicke
cosmology.

\section{spontaneous leptogenesis}

\hspace*{3.5mm}We consider a derivative coupling of the auxiliary
gravitational scalar $\phi$ to the ordinary matter. So we have now
an effective Lagrangian
 \begin {eqnarray}
 {\cal{L}}_{eff}=\frac{c}{M^2} \partial_{\mu} \phi J^{\mu},
 \end {eqnarray}
 where $M$ is the cutoff scale which, for example, could be Planck
 mass $M_{pl}$ or the scale of grand unification theory $M_{GUT}$,
 and $c$ is the coupling constant which characterizes the strength
 of gravitational scalar interacting with ordinary matter in the standard
 model of the electroweak theory. Taking $J^{\mu}=J^{\mu}_B$,
 during the evolution of the spatial flat FRW universe,
 ${\cal{L}}_{eff}$ in the equation (5) gives rise to an effective
 chemical potential $\mu_b$ for baryons:
 \begin {eqnarray}
\frac{c}{M^2} \partial_{\mu} \phi
J^{\mu}&\rightarrow&\frac{c}{M^2}
 \dot{\phi}n_B=\frac{c}{M^2}\dot{\phi}(n_b-n_{\bar{b}}), \\
\mu_b&=&c\frac{\dot{\phi}}{M^2}=-\mu_{\bar{b}}.
\end {eqnarray}
In thermal equilibrium the baryon number asymmetry is given by
(when $T\gg m_b$)
\begin {eqnarray}
n_B=\frac{g_b
T^3}{6}[\frac{\mu_b}{T}+{\cal{O}}(\frac{\mu_b}{T})^3]\simeq
c\frac{g_b \dot{\phi }T^2}{6M^2},
\end {eqnarray}
where $g_b$ counts the internal degrees of freedom of the baryon.
Using the familiar expression for entropy density
\begin {eqnarray}
s=\frac{2\pi^2}{45}g_{*}T^3,
\end {eqnarray}
we arrive at the final expression for the baryon to entropy ratio:
\begin {eqnarray}
\frac{n_B}{s}\simeq \frac{15c}{4\pi^2}\cdot\frac{g_b
\dot{\phi}}{g_{*}M^2T},
\end {eqnarray}
here $\dot{\phi}$ in Eq.(10) can be obtained by solving the
equation of motion of auxiliary scalar $\phi$ given below
\begin {eqnarray}
(2\omega-3)(\frac{\ddot{\phi}}{\phi}+3H\frac{\dot{\phi}}{\phi})+\frac{8\pi}{\phi}T_{m}=\frac{c}{M^2}[\dot{n_B}+(\dot{\phi}
+3H)n_B],
\end {eqnarray}
where $H$ is the Hubble constant and $T_{m}$ is the
energy-momentum scalar of the ordinary matter in the universe.

\hspace*{3.5mm}In applying the Brans-Dicke theory to cosmology, we
may write down the flat Robertson-Walker line element as
\begin {eqnarray}
ds^2=-dt^2+a^2(t)\tilde{g}_{ij}dx^idx^j.
\end {eqnarray}
Where i, j run from 1 to 3, $a(t)$ is the scale of the non-compact
3-dimensional flat space. The action of Brans-Dicke theory reads
\begin {eqnarray}
\nonumber I=&&\frac{1}{16\pi}\int d^4x\sqrt{-g}(\phi\cdot
R+\omega g^{\mu\nu}\frac{\nabla_{\mu}\phi\nabla_{\nu}\phi}{\phi})\\
&&+\int d^4x\sqrt{-g^4}\cdot{\textsc{L}}_{m},
\end {eqnarray}
here $\omega$ is the parameter of Brans-Dicke theory; Then the
corresponding field equation is given by
\begin {eqnarray}
 \nonumber
 R_{\mu\nu}-\frac{1}{2}g_{\mu\nu}R=&&-\frac{8\pi}{\phi}T_{\mu\nu}-
\frac{1}{\phi}(g_{\mu\nu}{\phi^{;\alpha}}_{;\alpha}-\phi_{;\mu;\nu})\\
 &&-\frac{\omega}{\phi^2}\phi_{;\mu}\phi_{;\nu}
+\frac{1}{2}\frac{\omega}{\phi^2}g_{\mu\nu}\nabla_{\sigma}\phi\nabla^{\sigma}\phi
\end {eqnarray}
The field equation for $\phi$ reads
\begin {eqnarray}
\Box^2\phi={\phi^{;\mu}}_{;\mu}=\frac{8\pi}{-3+2\omega}T_{m}.
\end {eqnarray}
Therefore, the fundamental equations of Brans-Dicke cosmology may
be taken as
\begin {eqnarray}
H^2&=&\frac{8\pi}{3}[\frac{\rho}{\phi}-\frac{\omega}{16\pi}\frac{\dot{\phi}^2}{\phi^2}
-(\frac{3H}{8\pi}\frac{\dot{\phi}}{\phi})];\\
\frac{\ddot{a}}{a}&=&-\frac{8\pi}{3}\frac{(-3+\omega)\rho+3\omega
p}{(-3+2\omega)\phi}+\frac{\omega}{3}\frac{\dot{\phi}^2}{\phi^2}+H\frac{\dot{\phi}}{\phi};\\
-\ddot{\phi}-3\frac{\dot{a}}{a}\dot{\phi}&=&\frac{8\pi}{-3+2\omega}(1-3w)\rho.
\end {eqnarray}
Here $w$ is the parameter of the state equation of the ordinary
matter($w=p/\rho$). The exact solutions to Brans-Dicke cosmologies
in flat Friedmann Universes had been drawn by R. Morganstern from
the early 70's \cite{Morganstern1,Morganstern2,Morganstern3}. He
found that as $a\rightarrow0$ the expansion parameter $a$ has the
same $t^{1/3}$ behavior for all curvature $k$ in radiation-filled
Brans-Dicke cosmology\cite{Morganstern2}. Moreover, as
$t\rightarrow0$ it is found that the flat-space solutions display
the same time dependence (and same dependence upon units) for $a$
and $\phi$ as do these same quantities in the radiation cases
($k=0,\pm1$)\cite{Morganstern3}. Therefore, we now turn to
investigate the feasibility of having spontaneous baryogenesis in
the early universe with Brans-Dicke gravity in the case of
$w=1/3$, so we have
\begin {eqnarray}
\rho\propto a^{-4}   \quad\quad and \quad\quad  \dot{\phi}\propto
a^{-3}.
\end {eqnarray}

In the research of Brans-Dicke cosmology, a power law form for
both the scale-factor and scalar field has ever been applied to
investigate the dynamics of a self-interacting Brans-Dicke field
to account for the acceleration of the universe at late epochs
$t\rightarrow t_0$ in \cite{Bertolami}. More importantly, the
exact general solutions to Brans-Dicke cosmology for $p=\rho/3$
found in \cite{Morganstern2} has also demonstrated a limiting
power-law behavior for early epochs $t\rightarrow0$. For these
reasons, we also consider a power-law form for the scale factor as
an approximation to find its asymptotic behavior in the early
universe,
\begin {eqnarray}
a=a_{in}(\frac{t}{t_{in}})^n.
\end {eqnarray}
Thus we have $H=n/t$, then Eq.(18) of $w=1/3$ immediately yields
the power law form for the scalar field
$\dot{\phi}=\dot{\phi}_{in}(t/t_{in})^{-3n}$ and
$\phi=\dot{\phi}_{in}(t/t_{in})^{1-3n}\cdot t_{in}\frac{1}{1-3n}$.
Giving this result, now we can write down the parameterized
Friedmann equation of (16) in a radiation-dominant epoch,
\begin {eqnarray}
n^{2}t^{-2}=\frac{8\pi(1-3n)\rho}{3\dot{\phi}_{in}t_{in}}\cdot(\frac{t_{in}}{t})^{1-3n}-n(1-3n)t^{-2}-\frac{\omega}{6}(1-3n)^{2}t^{-2}.
\end {eqnarray}

On the other hand, the radiation is in equilibrium so that
$\rho=\frac{\pi^2}{30}g_{*}T^4$ can be available in the final
calculation. Therefore, when the interaction of the spontaneous
baryogenesis is decoupled, a simple relation between the age and
temperature of the Universe is obtained,
\begin {eqnarray}
t^{-1-3n}&=&\frac{8\pi(1-3n)}{3\dot{\phi}_{in}[n^2+n(1-3n)+\omega(1-3n)^2/6]}\cdot
(t_{in})^{-3n}\frac{\pi^2}{30}g_{*}T^4.
\end {eqnarray}
Substituting Eq.(22) into the power law form for the scalar field,
we obtain
\begin {eqnarray}
\nonumber \dot{\phi}&=&\dot{\phi}_{in}(t/t_{in})^{-3n}\\
&=&\frac{8\pi(1-3n)}{3[n^2+n(1-3n)+\omega(1-3n)^2/6]}\cdot
t\frac{\pi^2}{30}g_{*}T^4 .
\end {eqnarray}
As far as the radiation-filled Brans-Dicke cosmology is concerned,
R. Morganstern has found the limiting behavior of the exact
solution for the expansion parameter is of $t^{1/3}$. Considered
spontaneous baryogenesis takes place at the early times of the
universe, when Eq.(21) possesses an asymptotic solution as
$n\rightarrow1/3$, so it requires that
\begin {eqnarray}
-\omega(1-3n)^2/6=n^2+n(1-3n).
\end {eqnarray}
in the limiting behavior. The equal sign of above equation would
come into existence only when $n=1/3$ in our model. However, to
say at least, Eq.(24) is still a fine approximation when we
consider the exact solution to radiation-filled Brans-Dicke
cosmology which possesses an asymptotic behavior of
$n\rightarrow1/3$ at the early epoches. Therefore, the ratio of
baryon number to entropy in this case is given by
\begin {eqnarray}
\frac{n_B}{s}\simeq
\frac{15c}{4\pi^2}\cdot\frac{8\pi^3g_b}{90M^2}\cdot\frac{(1-3n)}{n^2+n(1-3n)+\omega(1-3n)^2/6}\cdot
tT^3,
\end {eqnarray}
where $t$ and $T$ refer to the age and temperature of the
Universe, respectively.

 \hspace*{3.5mm}In the scenario of spontaneous baryogenesis, the baryon asymmetry is generated in thermal
equilibrium. This requires that baryon number violating
interactions occur rapidly ($\Gamma_{B}>H$). However, if the
$B$-violating interactions keep in equilibrium until
$\dot{\phi}\rightarrow0$, the final baryon asymmetry will be zero.
Denoting the epoch when the $B$-violating interactions freeze out
by $T_D$(corresponding to $t_D$), i.e., $\Gamma_{B}(t_D)=H(T_D)$,
the final baryon number asymmetry is obtained
\begin {eqnarray}
\frac{n_B}{s}
\simeq1c\cdot\frac{(1-3n)}{n^2+n(1-3n)+\omega(1-3n)^2/6}\cdot\frac{t_D{T_D}^3}{M^2}.
\end {eqnarray}
In the above numerical calculations, we have used $g_b\sim
{\cal{O}}(1)$. Furthermore, we may a priori assume that
$\frac{(1-3n)}{n^2+n(1-3n)+\omega(1-3n)^2/6}\sim{\cal{O}}(10^{-3})$
for considering $|\omega|>500$. However, as we shall see in the
last of this section, it will not change our result much if we
assume this expression takes the value in the order $\sim10^{-3}$.
After the interaction of the spontaneous baryogenesis decouples
the universe will continue to evolve into a radiation-dominated
epoch until the primordial nucleosynthesis takes place. But we
must note that thereafter the evolutional behavior of the scale
factor will deviate from the $n\rightarrow1/3$ law along with the
increasing of the time $t$. Therefore, we must be interested in
the epoches of the back direction from this decoupling time. As a
very good approximation, we can recall the relation of equilibrium
thermodynamics that $\frac{t_D}{t}=
(\frac{a_D}{a})^{4/4n}=(\frac{\rho_D}{\rho})^{-\frac{1}{4n}}=(\frac{\pi^2g_*/30T_D^4}{\pi^2g_*/30T^4})^{-\frac{1}{4n}}$
to obtain the evolution of the temperature. To make our discussion
to be as independent as possible of the cosmological evolution
models, here we choose the values of $T_{GUT}$ at the grand
unification phaser transitions epoch to be the input parameters in
our model. Firstly, we choose $T_{GUT}\sim10^{15} Gev$ considered
from particle physics. Then the induced baryon number asymmetry in
radiation-filled Brans-Dicke cosmology is
\begin {eqnarray}
\nonumber \frac{n_B}{s}&\simeq&
10^{-3}c\cdot\frac{t_{GUT}{T_{GUT}}^{\frac{1}{n}}{T_D}^{3-\frac{1}{n}}}{M^2}|_{n\rightarrow1/3}\\
&\simeq&1c\times10^{42}Gev^{3}\cdot\frac{t_{GUT}}{{M^2}_{pl}},
\end {eqnarray}
here we have taken the cut-off factor $M\simeq M_{pl}$. Then
yielding a sufficient $\frac{n_B}{s}\simeq10^{-10}$ would require
that $t_{GUT}\sim10^{-14}Gev^{-1}\sim10^{-38}sec$. This value is
not very far away from the Planck time $(t_{pl}\sim10^{-43}sec)$
but still can be achieved, just need the grand unification phase
transition to take place in a little advanced comparing with that
of Einstein's cosmology. If we consider that the index of the
power law $n$ will show a preference of a little departure from
$1/3$ along with the decrease of the temperature until $T_D$, then
the requirement on the time when the grand unification becomes to
take place would be further released to be $>10^{-38}sec$. Along
this way, the spontaneous baryogenesis using the Brans-Dicke
scalar is also possible to yield sufficient $n_B/s$ at a
radiation-dominated epoch.

On the other hand, if the B-violating interactions conserve
 $B-L$, the asymmetry generated will be erased by the eletroweak
 Sphaleron \cite{manton}. Hence, now we turn to leptogenesis \cite{fukugida,langacker}.
 We take $J^{\mu}$ in Eq.(5) to be ${J^{\mu}}_{B-L}$.
 Doing the calculations with the same procedure as above for
 $J^{\mu}={J^{\mu}}_{B}$ we have the final asymmetry of the baryon
 number minus lepton number:
 \begin {eqnarray}
\frac{n_{B-L}}{s}\simeq1c\cdot\frac{(1-3n)}{n^2+n(1-3n)+\omega(1-3n)^2/6}\cdot\frac{t_D{T_D}^3}{M^2}.
\end {eqnarray}
The asymmetry $n_{B-L}$ in (28) will be converted to baryon number
asymmetry when electroweak Sphaleron $B+L$ interaction is in
thermal equilibrium which happens for temperature in the range of
$10^2 GeV \sim 10^{12} GeV$. $T_D$ in (28) is the temperature
below which the $B-L$ interactions freeze out.

  \hspace*{3.5mm}In the Standard Model of the electroweak theory,
 $B-L$ symmetry is exactly conserved, however many models beyond
 the standard model, such as Left-Right symmetric model predict
 the violation of the $B-L$ symmetry. In this paper we take an
 effective Lagrangian approach and parameterize the $B-L$
 violation by higher dimensional operators. There are many
 operators which violate $B-L$ symmetry, however at dimension 5
 there is only one operator,
 \begin {eqnarray}
 {\cal{L}_{L}}=\frac{2}{f}l_Ll_l\chi\chi+H.c,
 \end {eqnarray}
 where $f$ is a scale of new physics beyond the Standard Model
 which generates the $B-L$ violations, $l_L$ and $\chi$ are the
 left-handed lepton and Higgs doublets respectively. when the
 Higgs field gets a vacuum expectation value $<\chi>\sim v$, the
 left-handed neutrino receives a majorana mass $m_{\nu}\sim
 \frac{v^2}{f}$.

\hspace*{3.5mm} In the early universe the lepton number violating
rate induced by the interaction in (29) is \cite{sarkar}
\begin {eqnarray}
\Gamma_{L}\sim0.04\frac{T^3}{f^2}.
\end {eqnarray}
Since $\Gamma_{L}$ is proportional to $T^3$, for a given $f$,
namely the neutrino mass, $B-L$ violation will be more efficient
at high temperature than at low temperature. Requiring this rate
be larger than the universe expansion rate $\sim n/t\sim
\frac{8n\pi(1-3n)}{3[n^2+n(1-3n)+\omega(1-3n)^2/6]}\cdot
\frac{1}{\dot{\phi}}\frac{\pi^2}{30}g_{*}T^4 $ until the
temperature $T_D$, we have
\begin {eqnarray}
f\sim(\frac{0.04}{n}t_D{T^3}_D)^{\frac{1}{2}}|_{n\rightarrow1/3}\sim(0.12t_D{T^3}_D)^{\frac{1}{2}}.
\end {eqnarray}
Therefore, we obtain a $T_D$-dependent lower limit on the neutrino
mass:
\begin {eqnarray}
\nonumber m_{\nu}&\sim&
\frac{v^2}{(0.12t_D{T^3}_D)^{\frac{1}{2}}}\sim\frac{6\times10^4Gev^2}{(0.12t_D{T^3}_D)^{\frac{1}{2}}}\\
&\sim& 1ev,
\end {eqnarray}
where we have using the energy scale of electroweak SSB phase
transition of about $246Gev$. Taking three neutrino masses to be
approximately degenerated, i.e., $m_1\sim m_2\sim m_3\sim
\tilde{m}$ and defining $\sum=3\tilde{m}$, one can see that for
$t_D{T^3}_D\sim 3\times10^{28}Gev^2$ GeV, three neutrinos are
expected to have masses $\tilde{m}$ around ${\cal{O}}(1 eV)$. The
current cosmological limit comes from WMAP \cite{spergel} and SDSS
\cite{tegmark}. The analysis of Ref.\cite{spergel} gives
$\Sigma<0.69 eV$. The analysis from SDSS shows, however that
$\Sigma<1.7 eV$ \cite{tegmark}. These limits on the neutrino
masses requires $t_D{T^3}_D$ be larger than
$5.7\times10^{29}Gev^2$ and $9.2\times10^{28}Gev^2$. The almost
degenerate neutrino masses required by the leptogenesis of this
model will induce a rate of the neutrinoless double beta decays
accessible for the experimental sensitivity in the near future
\cite{fujii}. Interestingly, a recent study \cite{allen} on the
cosmological data showed a preference for neutrinos with
degenerate masses in this range. Therefore, we may recall the
final expression (26) for spontaneous baryogenesis by using the
Brans-Dicke scalar. If the cut-off factor $M$ takes the value from
the grand unification scale $10^{16}$ GeV to the Planck scale
$10^{19}$ GeV, then yielding a sufficient baryon asymmetry may
requires that
$\frac{1}{3}\times10^{-6}<\frac{(1-3n)}{n^2+n(1-3n)+\omega(1-3n)^2/6}<1/3$.
It is quite easy to be achieved for a large number of Brans-Dicke
parameter $\omega$.

\hspace*{3.5mm}The experimental CPT test with a spin-polarized
torsion pendulum \cite{heckel} puts strong limits on the axial
vector background $b_{\mu}$ defined by
${\cal{L}}=b_{\mu}\bar{e}\gamma^{\mu}\gamma_5 e$ \cite{colladay}:
\begin {eqnarray}
|\vec{b}|\leq 10^{-28} GeV.
\end {eqnarray}
For the time component $b_0$, the bound is relaxed to be at the
level of $10^{-25}$ GeV \cite{Mocioiu}. In our model, assuming the
auxiliary scalar couples to the electron axial current the same as
Eq.(5), we can estimate the CPT-violation effect on the laboratory
experiments. As we know, the inverse of the Newtonian
gravitational constant $G$ takes the average value of the scalar
$\phi$, in the spirit of Brans-Dicke theory. The variation of $G$
in the present time is bounded as
$\frac{\dot{G}}{G}<1.2\times10^{-43}Gev$ by \cite{williams,
Dickey}. Hence the induced CPT-violating $b_0$ is
\begin {eqnarray}
\nonumber b_0&\sim&\frac{c}{M^2}\dot{\phi}_0\sim
\frac{c}{{M^2}_{gut}}\cdot\frac{-\dot{G}}{G^2}\\
&\leq&-1.2\times10^{-37} GeV,
\end {eqnarray}
which is much below the current experimental limits.

\section{gravitational leptogenesis}

\hspace*{3.5mm}In this section we will recall the gravitational
baryogenesis \cite{davoudiasl} in thermal equilibrium, which may
fails to work in the framework of Einstein's cosmology just has
been introduced in the beginning of this paper. So we investigate
it again in Brans-Dicke cosmology. The effective Lagrangian of a
$CP$-violating interaction between the derivative of the Ricci
scalar curvature $R$ and the baryon number(B) current $J^{\mu}$
still have the form of Eq.(2). Taking $J^{\mu}=J^{\mu}_B$,
 during the evolution of the spatial flat FRW universe,
 ${\cal{L}}_{eff}$ in the equation (2) gives rise to an effective
 chemical potential $\mu_b$ for baryons:
 \begin {eqnarray}
\frac{c}{M^2} \partial_{\mu} R J^{\mu}&\rightarrow&\frac{c}{M^2}
 \dot{R}n_B=\frac{c}{M^2}\dot{R}(n_b-n_{\bar{b}}), \\
\mu_b&=&c\frac{\dot{R}}{M^2}=-\mu_{\bar{b}}.
\end {eqnarray}
We could directly give out the final expression for the baryon to
entropy ratio in thermal equilibrium:
\begin {eqnarray}
\frac{n_B}{s}\simeq \frac{15c}{4\pi^2}\cdot\frac{g_b
\dot{R}}{g_{*}M^2T},
\end {eqnarray}
$\dot{R}$ in Eq.(37) can be obtained by solving the equation of
motion of the Ricci scalar given below
\begin {eqnarray}
\dot{R}&=&8\pi(\frac{\dot{T}_{m}}{\phi}-\frac{T_{m}\dot{\phi}}{{\phi}^2})+2\omega(\frac{\dot{\phi}\ddot{\phi}}{{\phi}^2}-\frac{{\dot{\phi}}^3}{{\phi}^3})-3(\frac{\dddot{\phi}}{\phi}-\frac{\dot{\phi}\ddot{\phi}}{{\phi}^2})
-9H(\frac{\ddot{\phi}}{\phi}-\frac{{\dot{\phi}}^2}{{\phi}^2})-9\dot{H}\frac{\dot{\phi}}{\phi}\\
&=&6\frac{\dddot{a}}{a}+6\frac{\ddot{a}\dot{a}}{a^2}-12\frac{\dot{a}^3}{a^3}.
\end {eqnarray}
Obviously, there are two methods which may be used to find the
behavior of $\dot{R}$ of the asymptotic solution
($n\rightarrow1/3$). The first simply involves substituting the
power law approximation of (20) in the function of the curvature
scalar (39) of the Robertson-Walker metric,
\begin {eqnarray}
\dot{R}=12n[(n-1)^2-n^2]\cdot t^{-3}.
\end {eqnarray}
The second method can be used even when the explicit solutions are
not known, and moreover it provides a convenient check on the
asymptotic behavior of the exact solution (namely the assumed
power law form (20) when $n\rightarrow1/3$). By substituting the
power law of the scale factor in Eq.(38)(which is resulted
directly from the field equation in Brans-Dicke cosmology (14)),
we again have
\begin {eqnarray}
\dot{R}=-2(1-3n)^2\omega \cdot t^{-3}.
\end {eqnarray}
On the other hand, here we should recall the relation (24) as a
fine approximation at the earliest times. After that, we can
compute the limiting behavior of $\dot{R}$ using both methods, and
of course we find that the results agree. Taking
$n\rightarrow1/3$, we immediately have $\dot{R}=12t^{-3}/9$. Thus
we obtain the ratio of baryon number to entropy, which is given by
\begin {eqnarray}
\frac{n_B}{s}\simeq \frac{15c}{4\pi^2}\cdot\frac{12
g_b}{9g_{*}M^2T}\cdot t^{-3},
\end {eqnarray}
where $t$ and $T$ also refer to the age and temperature of the
Universe. A rapidly violating of baryon number is still required
in this case. In addition, if the $B$-violating interactions keep
in equilibrium until $\dot{R}\rightarrow0$, the final baryon
asymmetry will be zero. Denoting the epoch when the $B$-violating
interactions freeze out by $T_D$(corresponding to $t_D$), i.e.,
$\Gamma_{B}(t_D)=H(T_D)$, the final baryon number asymmetry in
this case may be given by
\begin {eqnarray}
\nonumber \frac{n_B}{s}&\simeq &\frac{15c}{4\pi^2}\cdot\frac{12
g_b}{9g_{*}M^2T_D}\cdot {t_D}^{-3}\\
&\simeq&\frac{c}{2}\times10^{-2}\cdot\frac{{t_D}^{-3}}{M^2T_D}.
\end {eqnarray}
In the numerical calculations above, we have also used $g_b\sim
{\cal{O}}(1)$ and $g_{*}\sim {\cal{O}}(100)$. Recalling the
equilibrium thermodynamics $t_D\sim
t_{GUT}{(\frac{T_{GUT}}{T_D}})^{\frac{1}{n}}$ at a
radiation-dominated epoch $w=1/3$, then the final result will be
given under the consideration of the limiting behavior of the
exact solution in \cite{Morganstern2},
\begin {eqnarray}
\nonumber\frac{n_B}{s}&\simeq&
\frac{c}{2}\times10^{-2}\cdot {t^{-3}}_{GUT}\frac{(10^{15}Gev)^{-\frac{3}{n}}{T_D}^{\frac{3}{n}}}{M^2T_D}|_{n\rightarrow1/3}\\
&\sim&\frac{c}{2}\times10^{-137}Gev^{-9}\cdot
{t^{-3}}_{GUT}\cdot\frac{{T_D}^{8}}{M^2}.
\end {eqnarray}
Similarly, if the B-violating interactions conserve
 $B-L$, the asymmetry generated by gravitational CP-violation will
 also be erased by the eletroweak
 Sphaleron \cite{manton}. So we again have to turn to leptogenesis \cite{fukugida,langacker}.
 We take $J^{\mu}$ in Eq.(35) to be ${J^{\mu}}_{B-L}$.
 Doing the calculations with the same procedure as above for
 $J^{\mu}={J^{\mu}}_{B}$ we have the final asymmetry of the baryon
 number minus lepton number in a similar way to spontaneous
 leptogensis
 \begin {eqnarray}
\frac{n_{B-L}}{s}\simeq \frac{c}{2}\times10^{-137}Gev^{-9}\cdot
{t^{-3}}_{GUT}\cdot\frac{{T_D}^{8}}{M^2}.
\end {eqnarray}
Taking $c\sim{\cal{O}}(1)$, $n_{B-L}/s\sim10^{-10}$ and
$M\sim10^{16}GeV$, for the desired result is obtained when
$T_D<10^{15}Gev$, then the time $t_{GUT}$ when the grand
unification becomes to take place may be required to be
 \begin {eqnarray}
t_{GUT}<10^{-37}sec.
\end {eqnarray}
That is to say, the gravitational baryogenesis in Brans-Dicke
cosmology also requires the grand unification phase transition
take place in a little advanced comparing with that of Einstein's
cosmology.

 In the same way, we investigate the CPT-violation effect on the
laboratory experiments for the coupling between the curvature
scalar and the electron axial current. Firstly, viewed from Taylor
expansion, when a variable quantity evolves to a great number, the
derivative of a function of this quantity would tend to be more
and more smooth and flatted along with the increasing of the rank
of the derivative. Thus we have
$\frac{d}{dt}\ddot{a}\ll\frac{d}{dt}\dot{a}\ll\frac{d}{dt}a$ at
the present time(here the infinitesimal unit $\bigtriangleup t\sim
1sec$ is chosen to obtain a large number for the age of the
present universe relative to considered time interval). Secondly,
the Hubble's parameter at the present time has been given as
$H_0\sim10^{-17}sec^{-1}\sim2\times10^{-42}Gev$. Therefore the
induced CPT-violating $b_0$ in this case may be estimated at
\begin {eqnarray}
\nonumber b_0&\sim&\frac{c}{M^2}\dot{R}_0\sim
\frac{c}{M^2}[-6\frac{\ddot{a}\dot{a}}{a^2}+6\frac{\dddot{a}}{a}-12H^3+12\frac{\ddot{a}}{a}\cdot H]|_{t\rightarrow t_0}\\
\nonumber &\leq& \frac{c}{M^2_{gut}}[2.4\times10^{-83}Gev^2\cdot
sec^{-1}+1.2\times10^{-41}Gev\cdot
sec^{-2}-9.6\times10^{-125}Gev^3]\\
&\leq& 4.8\times10^{-122} GeV,
\end {eqnarray}
which also is much below the current experimental limits.

\section{conclusion}

\hspace*{3.5mm}In summary, we propose in this paper a scenario of
spontaneous baryogenesis in Brans-Dicke cosmology by introducing a
derivative coupling of the auxiliary scalar $\phi$ to the ordinary
matter. Our model is capable to explain the baryon number
asymmetry $n_{B}/s\sim10^{-10}$ without conflicting with
experimental tests on CPT. As a complementarity, we also
investigate the scenario of gravitational baryogenesis in
radiation-filled Brans-Dicke cosmology from a derivative coupling
of the Ricci scalar to the ordinary matter introduced by
Davoudiasl et al \cite{davoudiasl}, in this case the current
baryon asymmetry can also be achieved at early epoches just for a
non-vanishing $\dot{R}$, in different to the model in Einstein's
cosmology. As a result, we should also keep in mind, to obtain a
sufficient baryon asymmetry in both these two models in
Brans-Dicke cosmology, the time when the grand unification becomes
to take place would be required to be in a little advanced
comparing with that of Einstein's cosmology.

After this work was firstly submitted a related independent
work\cite{trodden04} appeared in the archives, which has similar
motivations\footnote{Recently a related paper appeared \cite{r05}.}.

\subsection*{Acknowledgment}
This work has been supported in part by the National Natural
Science Foundation of China under grant No. 10273017.

\end{document}